\documentclass{webofc}

\usepackage[varg]{txfonts}   
\usepackage{hyperref}
\usepackage{url}
\hypersetup{colorlinks=true,citecolor=blue,urlcolor=blue,linkcolor=blue}
\begin{document}
\title{First order transition region of an equation of state for QCD with a critical point}
%
%

\author{\firstname{Jamie} \lastname{Karthein}\inst{1}\fnsep\thanks{\email{jmkar@mit.edu}} \and
        \firstname{Volker} \lastname{Koch}\inst{2} \and
        \firstname{Claudia} \lastname{Ratti}\inst{3}
}

\institute{Center for Theoretical Physics -- a Leinweber Institute, Massachusetts Institute of Technology, Cambridge, MA 02139, USA 
\and
           Nuclear Science Division, Lawrence Berkeley National Laboratory, 1 Cyclotron Road, Berkeley, CA 94720, U.S.A. 
\and
           Physics Department, University of Houston, Houston TX 77204, USA
          }

\abstract{In addition to signals for the critical point, evidence for a first order phase transition would indicate a nontrivial structure within the QCD phase diagram. Moreover, while not a direct measurement of the critical point, the presence of a first order transition would imply its existence. This motivates the need to understand signatures of this first order transition in addition to directly studying the effect of a critical point. To this effect, we map the mean-field Ising model equation of state onto the QCD phase diagram, and reconstruct the full coexistence region in the case of a first order phase transition. Beyond the coexistence line, we maintain access to the spinodal region in the phase diagram, thus providing a description of metastable and unstable phases of matter. Thus, we describe the super-heated hadronic phase and the super-cooled quark-gluon plasma, which is useful for hydrodynamic simulations of the fireball created in a heavy-ion collision at low collision energy, where a first order phase transition is expected. We discuss the features of the pressure and other thermodynamic observables as functions of temperature and baryonic chemical potential, in particular their behavior in the coexistence region. Finally, we compare our equation of state to 3D Ising model ones available in the literature.
}
\maketitle
\section{Introduction}
\label{intro}

A key open question in the study of strongly interacting matter is whether a QCD critical point exists, separating a crossover at low density from a first-order phase transition at high baryon chemical potential. 
The Second Beam Energy Scan at RHIC aimed to explore this, with recent results on net-proton cumulants at low collision energies ($\sqrt{s_{NN}} \leq 20$ GeV) showing intriguing trends but no definitive evidence. 
Interpreting these findings requires an equation of state that incorporates tunable critical behavior, allowing exploration of different possible locations and strengths of the critical point \cite{Parotto:2018pwx,Karthein:2021nxe,Kapusta:2022pny,Kahangirwe:2024cny}.
If it turns out that nature has made the critical region small (as suggested by functional methods \cite{Fu:2021oaw}), it would be challenging to experimentally measure the critical point.
However, after the second-order phase transition (critical point) is the first-order phase transition line.
Thus, although not a direct measurement, the presence of a first order transition implies a preceding critical point.

An equation of state that describes this first order phase transition arises from Landau theory (also known as the mean field Ising model).
The mean field equation of state is determined by the dependence of the free energy on the magnetization ($M$), and temperature ($r=(T-T_c)/T_c$).
We may also define the external magnetic field $h=\partial F / \partial M|_r$.
Thus, according to Landau theory, we write the Helmholtz free energy, magnetic field, and pressure as:

\begin{align} 
    \centering
    F(M,r) &= h_0 \, M_0 \Big{(} \frac{1}{2}a \, r \, M^2 + \frac{1}{4}b \, M^4 \Big{)} \label{eq:MF_EoS_F}  \\
    h(M,r) &= \Bigg{(}\frac{dF}{dM}\Bigg{)}_r = h_0 (a \, r \, M + b \, M^3) \label{eq:MF_EoS_h} \\
    P(M,r) &= - G(M,r) = M \, h - F(M,r)   \label{eq:MF_EoS_P} \\ &= h_0 \, M_0 \Big{(}\frac{1}{2} a \, r \, M^2 + \frac{3}{4} b \, M^4 \Big{)} \nonumber,
\end{align}
With the choice  $a=3,~ b=1$, Landau theory reproduces a scaling equation of state \cite{Bzdak:2019pkr,Schofield:1969zza,Guida:1996ep} in terms of the parametric variables $(R,\theta)$, specifically the linear parametric model. 
We note here that by utilizing Landau theory, i.e. the mean field Ising model, we may forgo the formulation of the equation of state in terms of parametric variables and rely on the equations presented above which allow one to solve the system analytically with rational exponents.
Furthermore, by choosing the mean field Ising model, we maintain a complete description of the first order phase transition region of the phase diagram, including the coexistence and spinodal lines.
On the other hand, other approaches utilizing the 3D Ising model \cite{Parotto:2018pwx,Karthein:2021nxe,Kahangirwe:2024cny} are unable to describe the spinodals in the phase diagram due to the fact that they are pushed into the complex plane due to the non-trivial 3D critical exponents \cite{An:2017brc}.
Here, we focus on characterizing the first order region as an indirect method of detecting the  presence of a critical point in the QCD phase diagram.

\section{Spinodals in the QCD Phase Diagram}
\label{sec-1}

\begin{figure}
\centering
\includegraphics[width=7cm]{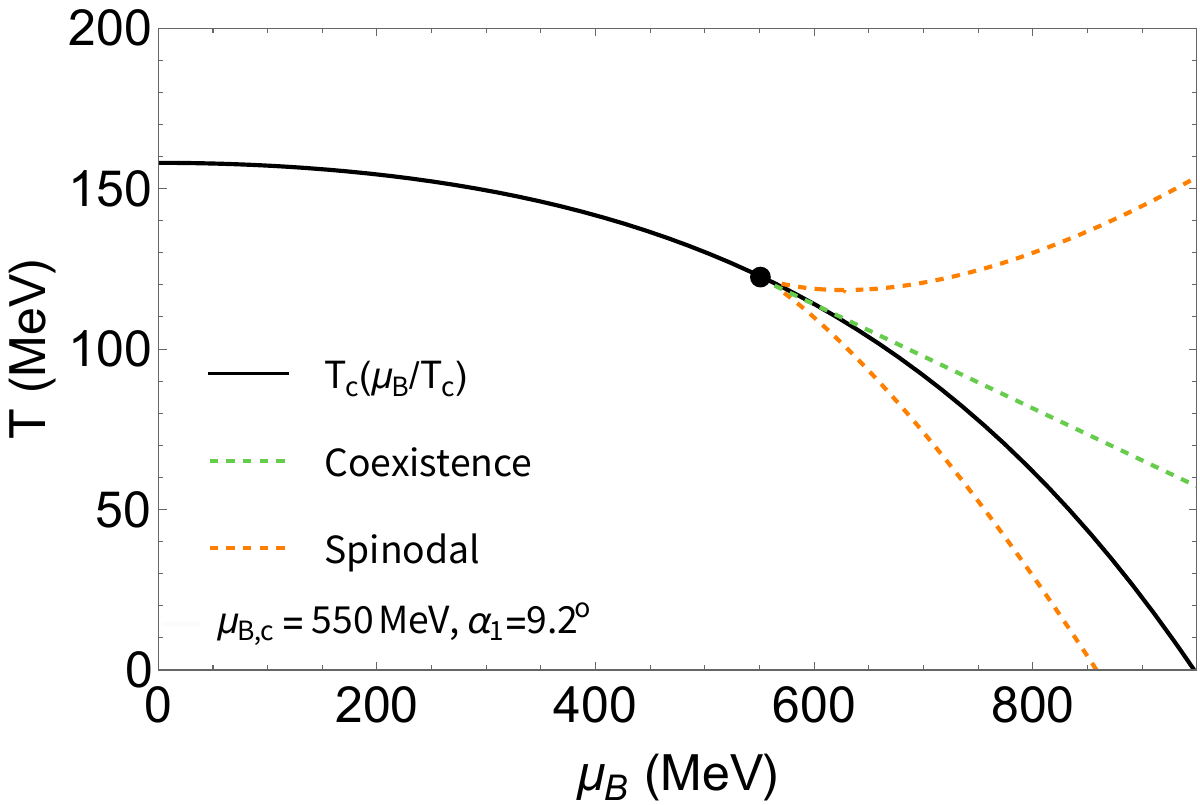}
\caption{The spinodal and coexistence curves on the QCD phase diagram as mapped from the mean field Ising model for a choice $\mu_{B,c}=550$ MeV.}
\label{fig:SpinodalsPhDiag_550}
\end{figure}

\begin{figure}
\centering
\includegraphics[width=7cm]{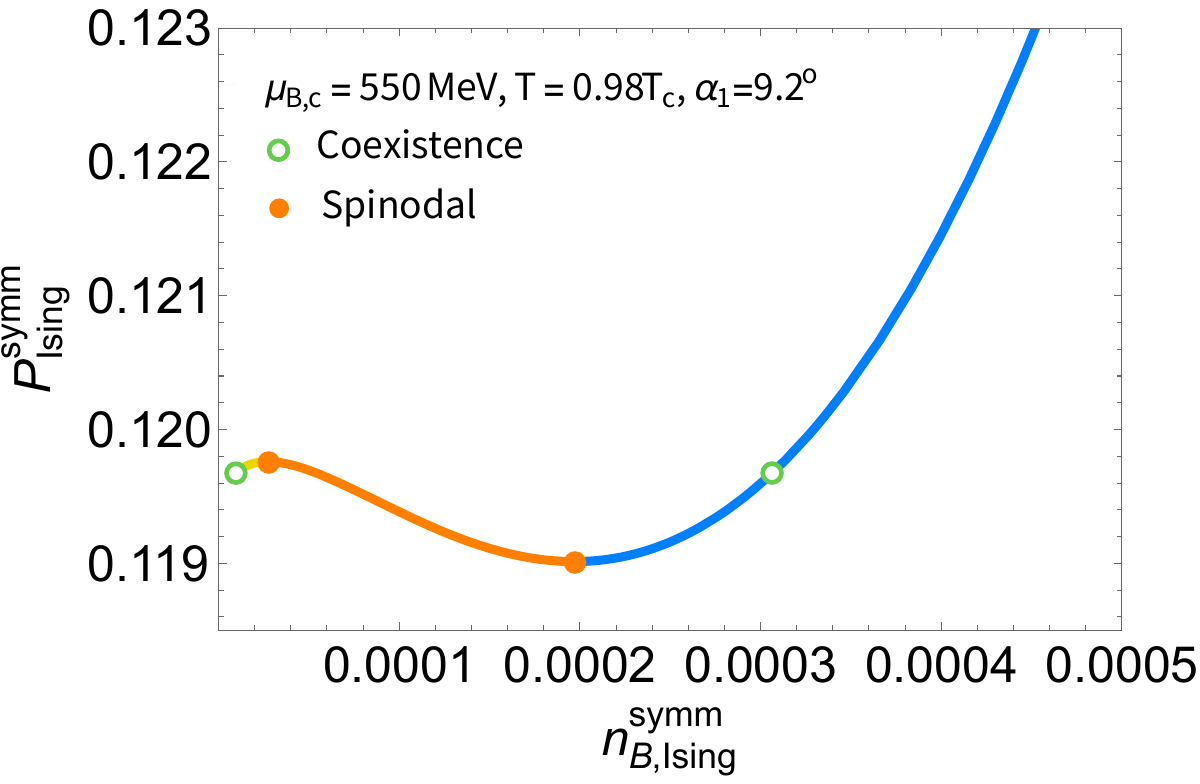}
\caption{An isothermal curve ($T=0.98 T_c$) for the Ising model pressure as a function of the density. The coexistence points are indicated with the open green circles, while the spinodal points are given by the filled orange circles.}
\label{fig:PvnB_550}
\end{figure}

In this work, we map the phase diagram from the mean field Ising model (in terms of reduced temperature $r=(T-T_c)/T_c$ and magnetic field $h$) onto the QCD one (in terms of temperature $T$ and baryonic chemical potential $\mu_B$).
By doing so, we introduce not only a critical point but also the first-order phase transition features on top of the equation of state from lattice QCD. 
When the transition is first order, a distinctive feature in the phase diagram is the spinodal region. 
The spinodal lines, or spinodal singularities, show the limits of metastability in the phase diagram \cite{Fisher:1967,Binder1987}.
In this region, thermodynamical quantities like entropy density, baryon density or energy density are multi-valued functions of the temperature at fixed chemical potential. 
Our main goal is to provide a full description of the equation of state around the first-order phase transition.
To this end, we implement the spinodal lines in the QCD phase diagram by mapping this feature from the mean field Ising model.

The mapping between the Ising model and QCD is not known from first-principles, as it is not a universal feature of the system, but rather depends on the microscopic details.
Thus, we utilize a linear mapping between Ising and QCD variables, as has been studied in detail in Refs. \cite{Parotto:2018pwx,Pradeep:2019ccv,Karthein:2021nxe, Mroczek:2022oga,Pradeep:2024cca,Karthein:2025hvl}:
    \begin{equation}
    \begin{split} \label{eq:mapTmuB}
        \frac{T-T_c}{T_c} &= \omega(\rho \, r \sin{\alpha_1} + h \sin{\alpha_2}) \\
        \frac{\mu_B-\mu_{B,c}}{T_c} &= \omega(-\rho \, r \cos{\alpha_1} - h \cos{\alpha_2})
    \end{split}
    \end{equation} 
where $T_c \rm{\,and\,} \mu_{B,c}$ are the coordinates of the critical point, and $\alpha_1 \rm{\,and\,} \alpha_2$ are the angles between the axes of the QCD phase diagram and the Ising model ones. 
Finally, $\omega$ and $\rho$ determine the contribution of the Ising phase diagram to the QCD phase diagram: $\omega$ is the overall scaling factor, while $\rho$ scales only the Ising transition line where $h=0$, or the $r$-axis.
This makes a total of 6 non-universal parameters in this map.
We reduce the number of free parameters from 6 to 4 by relying on constraints from  lattice QCD on the chiral crossover transition line, $T^{\rm{latt}}_{pc}(\mu_B)$ \cite{ Bonati:2015bha,Bellwied:2015rza,Bonati:2018nut,Bazavov:2018mes,Borsanyi:2020fev}.
Given a choice of $\mu_{B,c}$, this constraint determines the angle $\alpha_1$ and the critical temperature $T_c$.

We demonstrate the features of the coexistence and spinodal regions with an example choice of $\mu_{B,c}=550$ MeV, $w=1, \, \rho=2$ in Figs. \ref{fig:SpinodalsPhDiag_550} and \ref{fig:PvnB_550}.
For this choice of critical chemical potential, the metastable region occupies a larger part of the phase diagram as can be seen in Fig. \ref{fig:SpinodalsPhDiag_550}. 
In addition to the coexistence and spinodal lines, we also show the chiral phase transition line as a solid line with the critical point location given by the filled circle. 
In this case, the transition line ($h=0$) from the Ising model is a straight line tangent to the chiral phase transition line due to our choice of the linear map shown in Eq. \eqref{eq:mapTmuB}.
Ideally, the coexistence line should curve toward the $\mu_B$-axis, but since the focus of this work is on modeling the spinodal region, refining the mapping is left for future work.

We also show the behavior of a pressure-versus-density isotherm in Fig. \ref{fig:PvnB_550} after the mapping procedure and a symmetrization procedure that ensures the CP symmetry of QCD is obeyed.
Using the mean-field Ising approach, the full phase transition curve—including both coexistence points and spinodal points—can be mapped out. 
Three different solutions are shown (yellow, orange, and blue), with coexistence points marked by open circles and spinodal points by filled ones. 
The pressure vs. baryon density behavior confirms the presence of a mixed phase, with the orange curve indicating a region of mechanical instability ($\partial P/\partial n_B < 0$), characteristic of the unstable phase.

\section{Acknowledgments}
This material is also based upon work supported by the National Science Foundation under grants No. PHY-2514763, PHY-2208724, and PHY-2116686, and within the framework of the MUSES collaboration, under grant number No. OAC-2103680. This material is also based upon work supported by the U.S. Department of Energy, Office of Science, Office of Nuclear Physics, under Award Number
DE-SC0022023 and under contract number 
DE-AC02-05CH11231 as well as by the National Aeronautics and Space Agency (NASA) under Award Number 80NSSC24K0767.
J.M.K. is supported by an Ascending Postdoctoral Scholar Fellowship from the National Science Foundation under Award No. 2138063.

\bibliography{all.bib} 

\end{document}